\documentclass[
    ,final            
  ]
  {aipproc}

\newcommand{\pbp}{\bar{\psi}\psi}
\newcommand{\Tspace}{\rule{0pt}{2.6ex}}
\newcommand{\Bspace}{\rule[-1.2ex]{0pt}{0pt}}

\usepackage{amsmath}
\layoutstyle{8x11single}

\begin{document}

\title{The QCD Phase Transition Region\\ with Domain Wall Quarks}


\author{Zhongjie Lin}{
  address={Physics Department, Columbia University\\
  (HotQCD Collaboration)}
}

\begin{abstract}
  Results will be presented from a study of the QCD transition region 
  using $2+1$ flavors of fermions and a dislocation suppressing gauge action 
  on a lattice with temporal extent of 8 and spatial extent 16 ($1.9-2.7$ fm). 
  A series of temperatures from 140 through 200 MeV, separated by 10 MeV have been 
  studied. All the simulations lie on a line of constant physics with 200 MeV pions, 
  realized using domain wall fermion, a chirally symmetric fermion formulation. 
  The chiral condensates, susceptibility, anomalous symmetry breaking and a 
  detailed study of the Dirac spectrum will be described and compared with 
  earlier staggered results.
\end{abstract}

\maketitle


The chiral phase transition of strongly-interacting matter has 
been an intriguing topic for decades. Many significant efforts, 
both theoretical and experimental, have been dedicated to 
understanding this phenomena. Lattice Quantum Chromodynamics, 
since the day of its birth, has offered a powerful technique 
for an {\it ab initio}, non-perturbative study of the transition 
region.  Here we present a preliminary study of the transition 
region using a chiral formulation of lattice fermions in order 
to more accurately realize the chiral symmetry responsible for 
the transition.

\section{Gauge and Fermion actions}

While the gauge field can be implemented on the lattice in a 
straightforward manner, the usual lattice fermion schemes suffer
from problems such as breaking of chiral symmetry or creating 
doublers {\it etc.}, which interfere with the exploration of the 
QCD chiral phase transition using lattice methods. The domain wall 
fermion (DWF) formulation \cite{Kaplan:1992bt, Furman:1994ky}, a 
variant of Wilson fermions, avoids these problems at the 
cost of introducing an extra, $s$ dimension. 
The left and right-handed chiral states are bound to the four 
dimensional boundaries (walls) at the two ends of the 5-D volume.
Unphysical, massive degrees of freedom from the fifth dimension
are to a large extent canceled by the introduction of a Pauli-Villars 
term in the simulation. The low modes produce a discrete, four-dimensional 
version of QCD.

The DWF formulation has chiral breaking symmetry under good control.  The 
residual chiral symmetry breaking is characterized by a residual, additive 
quark mass: $ m_\mathrm{res}(L_s) = c_1\exp{(-\lambda_cL_s)}/L_s +c_2/L_s$.
If the extension in fifth dimension, $L_s$ is taken to infinity, one 
will obtain the overlap fermion formulation which exactly preserves 
chiral symmetry.  The infinite $L_s$ limit commutes 
with the continuum extrapolation, allowing us to choose an appropriate
$L_s$ to circumvent the heavy computational burden of overlap fermions
while keeping the chiral symmetry breaking minimal.
Another benefit of the DWF formalism is that $U(1)_A$ symmetry is 
only broken by axial anomaly in contrast to staggered fermions, where 
the anomalous symmetry is also broken by $a^2$ artifacts. 

In the simulation of QCD thermodynamics on a lattice with constant
physical quark masses, one naturally needs a small bare quark mass.
Since the residual mass is an additive correction to the bare quark 
mass in DWF formalism, {\it viz.} $m_q=m_\mathrm{res}+m_\mathrm{input}$,
we must also minimize $m_\mathrm{res}$.  From expression for $m_{\rm res}$, 
the residual mass can be decomposed into two parts. The first part is 
produced by standard five dimensional states and is exponentially suppressed; 
the other, comes from gauge field dislocations associated with changing
topology and is only suppressed by $1/L_s$. Therefore, naively increasing 
$L_s$ will not only be computationally expensive but also inefficient.  
Fortunately, the topology change associated with the second term also
leads to zero modes of an unphysical 4-dimensional Dirac operator, 
$D_W(-M_5)$ so we could multiply a factor ({\it e.g.} determinant of 
that Dirac operator) to suppress those configurations. Since such a 
factor will freeze topological tunneling, we add a twisted mass term 
to the determinant and introduce the ratio:
\begin{equation}
  \label{eqn:dsdr}
  \mathcal{W}(M_0, \epsilon_b, \epsilon_f) =  
  \frac{\det\left[D^\dagger_W(-M_0)D_W(-M_0) + \epsilon^2_f\right]}
  {\det\left[D^\dagger_W(-M_0)D_W(-M_0) + \epsilon^2_b\right]}.
\end{equation}
This is the dislocation suppression determinant ratio (DSDR) 
\cite{Vranas:1999rz,Fukaya:2006vs,Renfrew:2009wu}.  This alters
predominately the ultraviolet part of the theory, so the physical 
quantities in which we are interested are minimally affected.

\begin{table}[ht]
  \centering
  \begin{tabular}{ccc|ccc|ccc} \hline
    $T\,(\mathrm{MeV})$&$\beta$&$L_s$&$m_{\mathrm{res}}a$
    &$m_la$&$m_sa$&
    $\Tspace \Bspace \left<\pbp\right>_l/T^3$&
    $\Delta\pbp/T^3$&$\chi_{l,disc.}/T^2$\\ \hline
    140&1.633&48&0.00612&-0.00136&0.0519&6.26(12)& 7.74(12)&36(3) \\ 
    150&1.671&48&0.00296& 0.00173&0.0500&6.32(29)& 6.10(29)&41(2) \\ 
    150&1.671&32&0.00648&-0.00189&0.0464&8.39(10)& 7.06(10)&44(3) \\ 
    160&1.707&32&0.00377&0.000551&0.0449&5.25(17)& 4.83(17)&43(4) \\
    170&1.740&32&0.00209&0.00175 &0.0427&4.03(18)& 2.78(18)&35(5) \\
    180&1.771&32&0.00132&0.00232 &0.0403&3.16(15)& 1.56(15)&25(4) \\
    190&1.801&32&0.00076&0.00258 &0.0379&2.44(9) & 0.71(9) &11(4) \\
    200&1.829&32&0.00046&0.00265 &0.0357&2.19(8) & 0.47(8) &10(3) \\
    \hline
  \end{tabular}
  \caption{Summary of the ensembles.}
  \label{tab:conf}
\end{table} 

Table \ref{tab:conf} summarizes the ensembles in our simulation. 
We use $2+1$ flavors of DWF with Iwasaki and DSDR gauge action. The
spatial extent of the lattice is 16 (about 1.9 - 2.7 fm).   The
temporal extent is fixed to be 8 and the $\beta$ values are chosen so
that we have a range of temperatures from 140 to 200 MeV. The bare 
quark masses are also adjusted to make our ensembles lie on a line 
of constant physics with a pion mass of about 200 MeV.

\section{Chiral Symmetry}

\begin{figure}[htb]
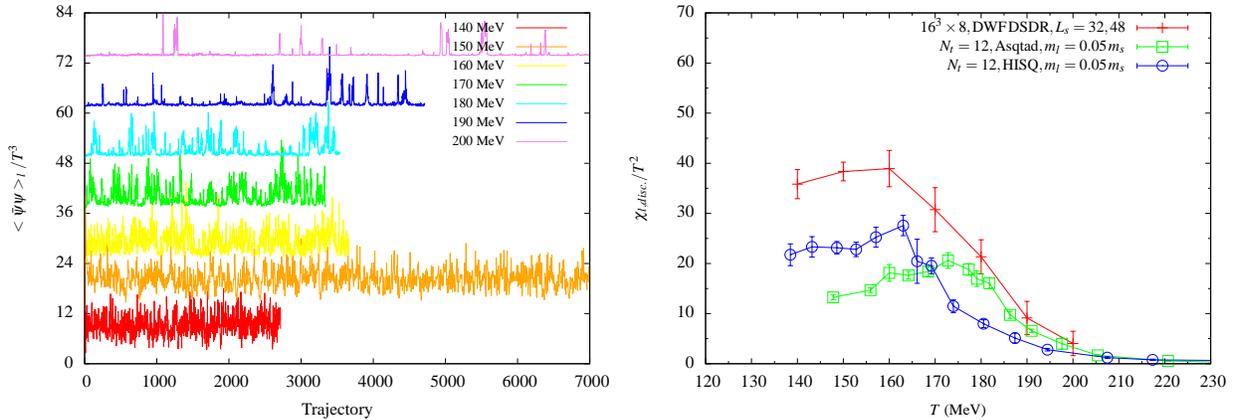

  \centering
  \begin{minipage}[t]{0.5\linewidth}
    \centering
    \resizebox{\linewidth}{!}{\input{./fig/Light.tex}}
  \end{minipage}%
  \begin{minipage}[t]{0.5\linewidth}
    \centering
    \resizebox{\linewidth}{!}{\input{./fig/Disc_Susc.tex}}
  \end{minipage}
  \label{fig:pbp}
  \caption{Evolution of $\left<\pbp\right>_l$, each shifted up by 
  12 units (left) and preliminary results of the disconnected 
  susceptibility, normalized in a consistent scheme differing from 
  that used in Table 1, at various temperatures (right).}
\end{figure}

The first quantity of interest is the chiral condensate shown as a
function of Monte Carlo time in fig. \ref{fig:pbp}.  As compared to 
those below 160MeV, the evolutions of the chiral condensate for 
temperatures above 160MeV display a distinct character, fluctuating 
above a base line.  The graph of the disconnected susceptibilities 
gives a more direct indication of a transition temperature around 
160 MeV, which agrees quite well with the staggered results.  However, 
since our spatial volume is quite limited (an aspect ratio 2), a 
quantitative comparison is likely premature. 

The chiral condensate is closely related to the small eigenvalues of the 
Dirac operator, which thus provides an alternative perspective into the 
symmetries associated with the QCD phase transition.  For instance, the 
well-known Banks-Casher relation \cite{Banks:1979yr}, 
$\lim_{m_l\to0}\lim_{V\to\infty} 
\langle\bar{\psi}_l\psi_l\rangle =-\pi\lim_{\lambda\to0}\lim_{m_l\to0}
\lim_{V\to\infty}\rho(\lambda)$, relates the chiral condensate and the
$\lambda=0$ value of the density $\rho(\lambda)$ of Dirac eigenvalues.

The Dirac spectrum also provides unique insights into the anomalous 
$U(1)_A$ symmetry.  Specifically, the $U(1)_A$-breaking difference of 
pseudoscalar and scalar susceptibilities is given by:
\begin{equation}
  \label{eqn:delta}
  \Delta_{\pi-\delta} \equiv \frac{\chi_{\pi} - \chi_{\delta}}{T^2}
  = \int\mathrm{d}\lambda\ \rho(\lambda)
  \frac{4m_l^2}{\left(m_l^2+\lambda^2\right)^2}.
\end{equation}
While $\Delta_{\pi-\delta}$ is expected to receive a small, dilute instanton
gas contribution from $\rho(\lambda) \propto m^2\delta(\lambda)$ at large
temperatures, there may also be larger, lower-temperature non-semi-classical 
contribution from $\rho(\lambda) \propto \lambda$.  We used Kalkreuter-Simma's 
method \cite{Kalkreuter:1995mm} to calculate the lowest 100 eigenvalues of the 
hermitian version of DWF Dirac operator.   The eigen-spectrum is then renormalized 
using an method similar to Giusti and Luscher \cite{Giusti:2008vb}.
\begin{figure}[htb]
  \centering
  \begin{minipage}[t]{0.33\linewidth}
    \centering
    \resizebox{\linewidth}{!}{\input{./fig/150MeV_1_norm.tex}}
  \end{minipage}%
  \begin{minipage}[t]{0.33\linewidth}
    \centering
    \resizebox{\linewidth}{!}{\input{./fig/160MeV_norm.tex}}
  \end{minipage}
  \begin{minipage}[t]{0.33\linewidth}
    \centering
    \resizebox{\linewidth}{!}{\input{./fig/170MeV_norm.tex}}
  \end{minipage} 
\end{figure}
\vskip -0.2in 
\begin{figure}[h]
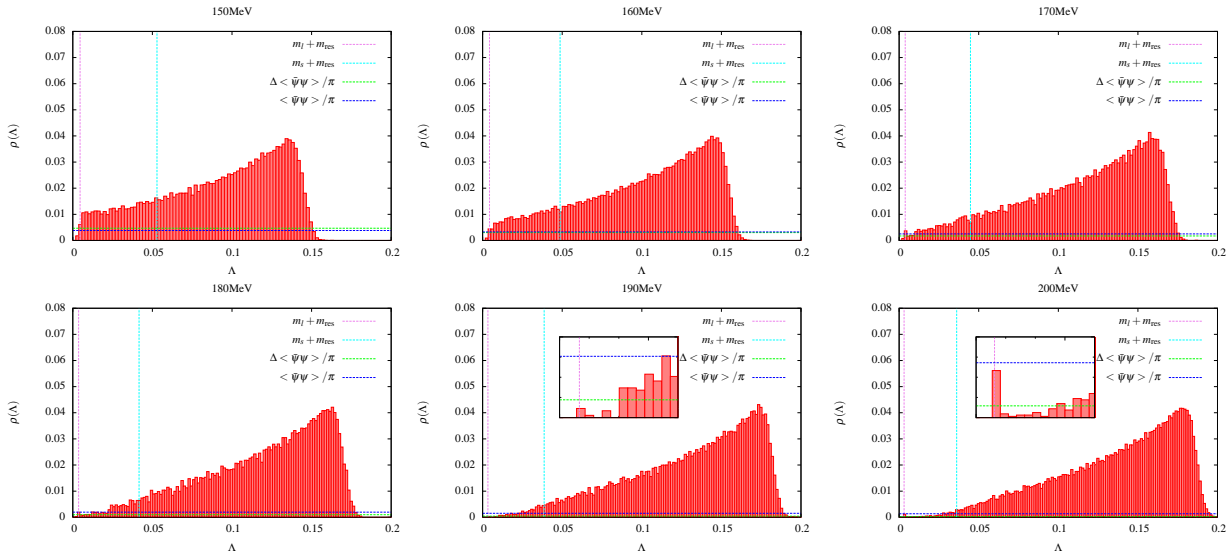


  \begin{minipage}[t]{0.33\linewidth}
    \centering
    \resizebox{\linewidth}{!}{\input{./fig/180MeV_norm.tex}}
  \end{minipage}%
  \begin{minipage}[t]{0.33\linewidth}
    \centering
    \resizebox{\linewidth}{!}{\input{./fig/190MeV_norm.tex}}
  \end{minipage}
  \begin{minipage}[t]{0.33\linewidth}
    \centering
    \resizebox{\linewidth}{!}{\input{./fig/200MeV_1_norm.tex}}
  \end{minipage}
  \label{fig:eig2}
  \caption{Dirac spectrum of $T=150 - 200$ MeV ensembles.}
\end{figure}

As can be seen in fig. \ref{fig:eig2}, the 
renormalization aligns the light quark mass (dashed vertical line) 
with the near-zero mode peak. It also compensates for the discrepancy 
in the Banks-Casher relation, which however still does not agree at 
about $50\%$ level below the transition. We attribute the discrepancy 
to the finite volume or finite mass effects. 
Above 170MeV, the chiral condensate and eigendensity both start to vanish
as expected for temperatures above the transition. But it remains
unclear whether the slope of the eigendensity vanishes at 180MeV. At
even higher temperature, a possible gap starts to emerge, indicating an 
effective restoration of $U(1)_A$ symmetry. 

We conclude that DWF provides a well-suited tool to study the QCD phase 
transition.  From the chiral susceptibility, we find indication of a 
transition temperature around 160MeV.  Well above the transition, we  
observe possible signals of $U(1)_A$ restoration.  Larger volume results 
and chiral extrapolation will be needed for a more definite conclusion. 


  I appreciate help and advice from N. Christ, F. Karsch and 
  R. Mawhinney and thank M. Cheng, P. Hegde and all members 
  of HotQCD as well as Columbia colleagues H. Yin and 
  Q. Liu for help and discussions.  This work was supported in part 
  by U.S. DOE grant DE-FG02-92ER40699. 
  The simulations were carried out on the BG/P machine at LLNL, the DOE- and 
  RIKEN-funded QCDOC machines and NYBlue machine at Brookhaven National Lab.



\bibliographystyle{aipproc}   

\bibliography{panic_proc}

\IfFileExists{\jobname.bbl}{}
 {\typeout{}
  \typeout{******************************************}
  \typeout{** Please run "bibtex \jobname" to optain}
  \typeout{** the bibliography and then re-run LaTeX}
  \typeout{** twice to fix the references!}
  \typeout{******************************************}
  \typeout{}
 }

\end{document}